\documentclass[showpacs,preprintnumbers,amsmath,amssymb,,nofootinbib]{revtex4}
\usepackage{epsfig}

\usepackage{graphicx}
\usepackage{epsfig}
\usepackage{bm}

\makeatletter \leftline{\epsfbox{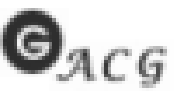}}
 \leftline
{\footnotesize{\textbf{\textsf{GACG-04/02}}}}
\begin{document}
\title{ Hamiltonian treatment of Collapsing Thin
Shells in Lanczos-Lovelock's theories}

\author{Juan Cris\'{o}stomo}
\altaffiliation{jcrisost@gacg.cl}
\author{Sergio del Campo}
\altaffiliation{sdelcamp@ucv.cl}
\author{Joel Saavedra}
\altaffiliation{joelium@gacg.cl} \affiliation{Instituto de
F\'{\i}sica, Facultad de Ciencias B\'{a}sicas y Matem\'{a}ticas,
Pontificia Universidad Cat\'{o}lica de Valpara\'{\i}so, Avenida
Brasil 2950, Valpara\'{\i}so, Chile.}

\begin{abstract}
The Hamiltonian treatment for the collapse of thin shells for a
family of Lanczos-Lovelock theories is studied. This formalism
allows us to carry out a concise analysis of these theories. It is
found that the black holes solution can be created by collapsing a
thin shell. Naked singularities cannot be formed by this
mechanism. Among the different Lanczos-Lovelock's theories,  the
Chern-Simons' theory corresponds to an exceptional case, because
naked singularities can emerge from the collapse of a thin shell.
This kind of theory does not possess a gravitational
self-interaction analogous to the Newtonian case.
\end{abstract}
\maketitle

\section{Introduction}

In a recent work \cite{ctz}, black hole solutions in a particular
class of Lovelock's gravitation theories were studied. These
theories were selected by requiring that they have a unique
Anti-de Sitter (AdS) vacuum with a fixed cosmological constant.
This strongly restricts the coefficients in the Lanczos-Lovelock
(\textbf{LL}) action \cite{LL}. For a given dimension $d$, the
Lagrangians under consideration are labelled by an integer
$k=1,...,\left[\frac{d-1}{2}\right]$\footnote{Here $\left[
x\right] $ is the integer of $x$.}, where the Einstein-Hilbert
(\textbf{EH}) Lagrangian corresponds to the case $k=1$. For
$k=\left[ \frac{d-1}{2}\right] $, we must distinguish between even
and odd dimensions, because the theories are different. When $d$
is odd, the corresponding Lagrangian is given by the
Euler-Chern-Simons form (\textbf{CS}) for the AdS group, whose
exterior derivative is proportional to the Euler density in $2n$
dimensions \cite{CHAM,CS}. For $d$ even, the Lagrangian reads as
the Born-Infeld form (\textbf{BI}). In this case the expression
for the (\textbf{LL}) action is proportional to the Pfaffian of
the $2$-form $\bar{R}^{ab}$ and, in this sense, it has a
Born-Infeld-like form \cite{BI}. These two cases are exceptional,
because they are the only ones which allow sectors with
non-trivial torsion \cite{HDG}. For $d\geq 7$ there exist other
interesting possibilities, which are different from \textbf{EH},
\textbf{BI} and \textbf{CS}. For example, the theory with $k=2$
has been studied by several authors in different scenarios
\cite{az,oz,el,pa}.

In this \textbf{LL} theories for any dimensions and $k$, there
exist well-behaved black hole solutions. However, we must
differentiate between cases with odd and even $k$, because in
theories with even $k$ an, additional solution appears, which
represents a naked singularity.

It is interesting to study the black holes formations through
gravitational collapses of thin shells. In the usual thin shell
treatment \cite {IS,KU,IP}, the analysis of collapse is based on
the discontinuities of the extrinsic curvature of the world tube
of the collapsing matter. However, the implementation of the
Israel formalism in this kind of theories (\textbf{LL}) is very
difficult, because the action contains high powers in the
curvature and, therefore, in the extrinsic curvature. In this
formalism the complicated analysis of collapse makes the treatment
quite unattractive.

Another approach in studying matter collapses is the
Oppenheimer-Snyder formalism, which was applied by Ilha  et al
\cite{lemos1,lemos2} to the case of a homogeneous collapsing dust,
where the inner metric is described by the
Friedmann-Robertson-Walker line element, and the external metric
corresponds to the solution of the fields equations in \textbf{BI}
or \textbf{CS} theories. In Ref. \cite{lemos2}, the authors
discussed the formation of a naked singulary in the \textbf{CS}
theory.

On the other hand, an alternative way to study gravitational
collapse of thin shell is the Hamiltonian treatment. This
treatment was applied in Ref. \cite{cot} to the Einstein-Hilbert
gravity, where the direct integration of the canonical constraints
reproduces the standard shell dynamics for a number of known
cases. In particular, it was applied in detail to three
dimensional spacetime and the properties of the (2+1)-dimensional
charged black hole collapse was further elucidated. The
Hamiltonian treatement was also extended to deal with rotating
solutions in three dimensions. The general form of the equations
of the shell dynamics implies the stability of black holes. As far
as, black hole in this model cannot be converted into naked
singularities by any shell collapse processes.

In this work we will extend the Hamiltonian formalism to our
approach to the theory of \textbf{LL} in high dimensions, and
particularly to the theories described in Ref. \cite{ctz}. This
formalism permits us to analyze the black holes formations in an
economical way.

The plan of the paper is as follows. Section II briefly reviews
the \textbf{LL} action and its spherically symmetric solution.
Section III analyzes the collapse of a spherically symmetric shell
under the Hamiltonian formalism. Finally, section IV is devoted to
conclusions.

\section{LL Action}
The \textbf{LL} action is a polynomial of degree $\left[
d/2\right] $ in the curvature, which can be also written in terms
of the Riemann curvature $R^{ab}=d\omega ^{ab}+\omega
_{c}^{a}\omega ^{cb}$ and the vielbein $ e^{a}$ as\footnote{ Wedge
product between forms is understood throughout.}

\begin{equation}
I_{(g)}=\kappa \int \sum\limits_{p=0}^{\left[ d/2\right] }\alpha
_{p}\epsilon
_{a_{1}...a_{d}}R^{a_{1}a_{2}}.....R^{a_{2p-1}a_{2p}}e^{a_{2p+1}}...e^{a_{d}}%
\text{,}  \label{s1}
\end{equation}
where $\alpha _{p}$ are arbitrary constants. In the first-order
formalism, the action (\ref{s1}) is regarded as a functional of
the vielbein and the spin connection, and the corresponding field
equations obtained varying with respect to $e^{a}$ and $\omega
^{ab}$ reads
\begin{equation}
\sum_{p=0}^{\left[ \frac{d-1}{2}\right] }\left( d-2p\right) \alpha
_{p}\epsilon
_{ab_{1}...b_{d-1}}R^{b_{1}b_{2}}...R^{b_{2p-1}b_{2p}}e^{b_{2p+1}}...e^{b_{d-1}}=0%
\text{,}  \label{sh1}
\end{equation}

\begin{equation}
\sum_{p=0}^{\left[ \frac{d-1}{2}\right] }p\left( d-2p\right) \alpha
_{p}\epsilon
_{abc_{3}...c_{d}}R^{c_{3}c_{4}}...R^{c_{2p-1}c_{2p}}T^{c_{2p+1}}e^{c_{2p+2}}...e^{c_{d}}=0%
\text{.}  \label{sh2}
\end{equation}
Here $T^{a}=de^{a}+\omega _{b}^{a}e^{b}$ is the torsion 2-form.

Note that in even dimensions, the term $L^{(d/2)}$ is the Euler
density and, therefore, does not contribute to the field
equations. However, the presence of this term in the action --with
a fixed weight factor-- guarantees the existence of a well-defined
variational principle for asymptotically locally AdS spacetimes
\cite{ACOTZ3+1,ACOTZ2n}.

The first two terms in the \textbf{LL} action (\ref{s1}) are the
cosmological and kinetic terms of the \textbf{EH} action.
Therefore, General Relativity is contained in the \textbf{LL}
theory as a particular case.

The linearized approximation of the \textbf{LL} and \textbf{EH}
actions around a flat, torsionless background are classically
equivalent \cite{Zumino}. However, beyond the perturbation theory
the presence of higher powers of curvature in the Lagrangian make
both theories radically different. In particular, black holes and
big-bang solutions of (\ref{sh1}) have different asymptotic
behaviors from their \textbf{EH} counterparts. Hence, a generic
solution of the \textbf{LL} action cannot be approximated by a
solution of Einstein's theory.

\subsection{Static and Spherically Symmetric Solutions}

Consider static and spherically symmetric solutions of equations (\ref{sh1})
and (\ref{sh2}). In Schwarzschild-like coordinates, the metric can be
written as

\begin{equation}
ds^{2}=-N^{2}(r)f^{2}(r)dt^{2}+\frac{dr^{2}}{f^{2}(r)}+r^{2}d\Omega
_{d-2}^{2}\text{.}  \label{s2}
\end{equation}
Replacing this Ansatz in the field equations (\ref{sh1}) and
(\ref{sh2}) leads to the following equations for $N(r)$ and
$f^{2}(r)$ \footnote{In the first order formalism, the field
equations imply that the torsion vanishes, except for \textbf{BI}
and \textbf{CS} theories, so that, it is not necessarily to set
$T^{a}=0$ \cite{HDG}. However, for static and spherically
symmetric configurations the equation (\ref{sh2}) implies that the
torsion must vanish in these cases as well.}

\begin{eqnarray}
\frac{dN}{dr} &=&0\text{,}  \label{ninf} \\
\frac{d}{dr}\left( r^{d-1}\sum_{p}(d-2p)\alpha _{p}\left( \frac{1-f^{2}}{%
r^{2}}\right) ^{p}\right) &=&0\text{.}  \label{sh3}
\end{eqnarray}
Integrating equations (\ref{ninf}) and (\ref{sh3}) yields

\begin{eqnarray}
N &=&N_{\infty }\text{,}  \label{n} \\
\sum_{p}(d-2p)\alpha _{p}\left( \frac{1-f^{2}}{r^{2}}\right) ^{p} &=&\frac{1%
}{\kappa (d-2)!\Omega _{d-2}}\frac{M+C_{0}}{r^{d-1}}\text{,}  \label{she1}
\end{eqnarray}
where the constant of integration $N_{\infty }$ relates the
coordinate time to the proper time of an observer at spatial
infinity. We will assume it equal to one. The constant $M$ stands
for the mass up to an additive constant $C_{0}$, which is nonzero
only in the case of \textbf{CS} theories \cite{ctz}.

Equations (\ref{she1}) corresponds to the solution of field
equations, which is a polynomial in $f^{2}(r)$, so many roots for
$f^{2}(r)$ with the same mass will exist, but, with different
asymptotical behavior. This means that (\ref{sh1}) possesses, in
general, several solutions with a constant curvature in the
asymptotical region, making the value of the cosmological constant
ambiguous. In fact, the cosmological constant could change in
different regions of the same spatial section, or it could jump
arbitrarily as the system evolves in time \cite{TH}.

These problems are overcome by demanding that the theory have a
unique cosmological constant \cite{ctz}. In order to satisfy this
condition, we choose the coefficients $\alpha _{p}$'s as follows:

\begin{equation}
\alpha _{p}:=c_{p}^{k}=\left\{
\begin{array}{l}
\frac{l^{2(p-k)}}{(d-2p)}\left(
\begin{array}{l}
k \\
p
\end{array}
\right) \text{, }p\leq k \\
0\text{ \qquad \qquad \quad , }p>k
\end{array}
\right. \text{,}  \label{coe}
\end{equation}
and

\smallskip
\begin{equation}
\kappa =\frac{1}{2(d-2)!\Omega _{d-2}G_{k}}\text{.}  \label{kappa}
\end{equation}

With this choice, $f^{2}(r)$ adopts the following form:

\begin{equation}
f^{2}(r)=\frac{r^{2}}{l^{2}}+1-\chi \left( \frac{2G_{k}M+\delta _{d-2k,1}}{%
r^{d-2k-1}}\right) ^{1/k}\text{,}  \label{f2}
\end{equation}
where $\chi =(\pm 1)^{k+1}$. For even $k$, the ambiguity of sign
expressed through $\chi $ in (\ref{f2}) implies that there are two
possible solutions, provided $M>0$. The solution with $\chi =1$
describes a black hole with an events horizon surrounding the
singularity at the origin. The solution with $ \chi =-1$ has a
naked singularity with positive mass. If $k$ is odd, there is no
ambiguity of sign because $\chi =1$. Therefore this solution
corresponds to a black hole with positive mass.

From eq. (\ref{f2}), it is observed that for $k=1$, the AdS black
hole solution for \textbf{EH} in $d$-dimensional is recovered. The
black hole solutions corresponding to \textbf{BI} and \textbf{CS}
theories \cite{CS} are obtained also from expression (\ref{f2}),
setting $k=\left[ \frac{d-1}{2}\right]$.

\section{Collapse of Thin Shells}

Let $\Sigma _{\xi }$ be a time-like hypersurface, which represents
the evolution of a thin shell \cite{IS,KU}. This hypersurface
divides the spacetime into two regions; the interior denoted by
$V^{(-)}$ and the exterior denoted by $V^{(+)}$, respectively.
Each of these regions contains $\Sigma _{\xi }$ as a part of its
boundary. We introduce into $\Sigma _{\xi }$ a set of intrinsic
coordinates $\rho ^{a}$, where the Latin indices go from $0$ to
$d-2$, and in the regions $V^{(-)}$ and $V^{(+)}$, the independent
coordinates $x_{-}^{\alpha }$ and $x_{+}^{\alpha }$ are
introduced, so that the parametric equations for $\Sigma _{\xi }$
in these charts are $x_{-}^{\alpha }(\rho ^{a})$ and
$x_{+}^{\alpha }(\rho ^{a})$, respectively.

At each point on $\Sigma _{\xi }$ there exists a unit space-like
vector $\xi ^{\mu }$, normal to $\Sigma _{\xi }$ and pointing from
$V^{(-)}$ to $V^{(+)}$, and $d-1$ vectors $e_{a}^{\alpha
}=\partial x^{\alpha }/\partial \rho ^{a}$ tangential to $\Sigma
_{\xi }$ in the directions of the coordinates $\rho ^{a}$.

The time-like hypersurface $\Sigma _{\xi }$ represents the
evolution of an infinitesimal $d-2$-dimensional matter thin shell.
There is no matter outside the shell. Therefore, the matter moves
only on the shell, so that its $d$-velocity $u^{\alpha }$ is
normal to $\xi ^{\lambda }$ and vanishes outside $\Sigma _{\xi }$.
Moreover, an observer on the shell can refer the movement of
matter to the reference points $(\rho ^{1},...,\rho ^{d-2})$ and
the reference time $\rho ^{0}=\tau $, and thus, the velocity is
described by an intrinsic vector $u^{a}$. The vectors $u^{\alpha
}$ and $u^{a}$ are joined by the relation
$u^{\alpha}=e_{a}^{\alpha }u^{a}$.

The mechanical properties of matters are described by the surface
energy-momentum tensor $T_{\mu \nu }$, which is normal to $\xi
^{\lambda }$ and it vanishes outside $\Sigma _{\xi }$. For an
observer on of $\Sigma _{\xi }$, the tensor $T_{\mu \nu }$ is
described by the intrinsic coordinates. For an ideal fluid, the
intrinsic energy-momentum tensor has this form:

\begin{equation}
T_{ab}=\sigma u_{a}u_{b}-\hat{\tau}(h_{ab}+u_{a}u_{b})\text{,}  \label{tab}
\end{equation}
where $\sigma $ means the rest surface mass density of the shell
and $ \hat{\tau}$ the surface tension\footnote{Here $\hat{\tau}$
denotes the surface tension and $\tau$, the proper time.}. Since,
the tensor $T_{ab}$ is confined into the hypersurface $\Sigma
_{\xi }$, it satisfies the continuity equation $T_{a/b}^{b}=0$.
Multiplying this tensor by $u^{a}$, we obtain the following
relation

\begin{equation}
(\sigma u^{a})_{/a}-\hat{\tau}u^{a}_{/a}=0\mbox{.}\label{state}
\end{equation}
This equation can be seen as the equation of state of the matter
on the hypersurface $\Sigma _{\xi }$.

The next step is to introduce a timelike ADM foliation,
$\Sigma_{t}$, of the spacetime. The foliation intersects the world
tube of the collapsing matter, which corresponds to the thin shell
of the $\Sigma_{t}$ at the time $t$. As usual, the metric tensor
is decomposed by using the basis $N^{\mu }=(N^{\bot },N^{i})$,
where $N^{\bot }$ represents the lapse and $N^{i}$ the shift
function. In this basis the line element in the coordinates
$x^{\alpha }$ of the regions $V^{(-)}$ and $V^{(+)}$ takes the
form

\begin{equation}
ds^{2}=-(N^{\bot })^{2}dt^{2}+g_{ij}(N^{i}dt+dx^{i})(N^{j}dt+dx^{j})\text{.}
\label{dsadm}
\end{equation}

In the presence of matter, and since $N^{\mu }$ are Lagrange
Multipliers, the total Hamiltonian becomes

\begin{equation}
\mathcal{H}=N^{\bot }\mathcal{H}_{\bot }+N^{i}\mathcal{H}_{i}\text{,}
\label{h}
\end{equation}
where $\mathcal{H}_{\bot }$ and $\mathcal{H}_{i}$ are defined by

\begin{equation}
\mathcal{H}_{\bot }=\mathcal{H}_{\bot }^{(g)}+\mathcal{H}_{\bot
}^{(m)}\text{ ,}  \label{hp}
\end{equation}
\smallskip and
\begin{equation}
\mathcal{H}_{i}=\mathcal{H}_{i}^{(g)}+\mathcal{H}_{i}^{(m)}\text{,}
\label{hi}
\end{equation}
respectively. Here $\mathcal{H}_{\bot }^{(g)}$ and
$\mathcal{H}_{i}^{(g)}$ correspond to the Hamiltonian terms
related to the gravitational field of the \textbf{LL }action
\cite{TZ}, and are given by

\begin{equation}
\mathcal{H}_{\bot }^{(g)}=-\kappa \sqrt{g}\sum\limits_{p}\frac{(d-2p)!\alpha
_{p}}{2^{p}}\delta _{j_{1}...j_{2p}}^{i_{1}...i_{2p}}R_{i_{1}i_{2}}^{\,\quad
j_{1}j_{2}}...R_{i_{2p-1}i_{2p}}^{\qquad j_{2p-1}j_{2p}}\text{,}  \label{hpg}
\end{equation}
and

\begin{equation}
\mathcal{H}_{i}^{(g)}=-2\pi _{i/j}^{j}\text{,}  \label{hig}
\end{equation}
where $\pi ^{ij}$ are the conjugate momenta to the metric tensor
of the intrinsic tensor metric $g_{ij}$; $g$, its determinant; and
$R_{ijkl}$ corresponds to the tensor curvature, which can be
written in terms of the geometric quantities of $\Sigma_{t}$ as
\begin{equation}
R_{ijkl}=\hat{R}_{ijkl}+K_{ik}K_{jl}-K_{il}K_{jk},
\end{equation}
with $\hat{R}_{ijkl}$ stands for the curvature tensor of the leaf
$\Sigma _{t}$ of the foliation and $K_{ij}$ is the extrinsic
curvature.

The momenta are defined in terms of extrinsic curvature $K_{ij}=\frac{1}{%
2N^{\bot }}(N_{i/j}+N_{j/i}-\dot{g}_{ij})$ as

\begin{equation}
\pi _{j}^{i}=-\kappa \sqrt{g}\sum\limits_{p}\frac{p!(d-2p)!\alpha _{p}}{%
2^{p+1}}\sum\limits_{s=0}^{p-1}D_{s(p)}\delta
_{jj_{1}...j_{2s}...j_{2p-1}}^{ii_{1}...i_{2s}...i_{2p-1}}R_{i_{1}i_{2}}^{\,%
\quad j_{1}j_{2}}...R_{i_{2s-1}i_{2s}}^{\qquad
j_{2s-1}j_{2s}}K_{i_{2s+1}}^{j_{2s+1}}...K_{i_{2p-1}}^{j_{2p-1}}\text{,}
\label{pi}
\end{equation}
where

\[
D_{s(p)}=\frac{(-4)^{p-s}}{s![2(p-s)-1]!!}\text{,}
\]

The matter components $\mathcal{H}_{\bot }^{(m)}$ and
$\mathcal{H}_{i}^{(m)}$, are given by

\begin{equation}
\mathcal{H}_{\bot }^{(m)}=\sqrt{g}T_{\bot \bot }\text{,}  \label{hpm}
\end{equation}

\begin{equation}
\mathcal{H}_{i}^{(m)}=2\sqrt{g}T_{\bot i}\text{,}  \label{him}
\end{equation}
where $\perp $ corresponds to a contraction with a normal vector
to the hypersurface $ \Sigma _{t}$, $n_{\mu }=(-N^{\perp
},0,...,0)$.

In what follows we restricts ourselves to the spherical
coordinates. We will use the proper time $\tau $ and spherical
angles as intrinsic coordinates of the hypersurface $\Sigma _{\xi
}$; $\rho ^{a}=(\tau ,\theta ^{1},...,\theta ^{d-2})$. The motion
of the shell is expressed by the equation $r=R(\tau )$. The
derivative with respect to $\tau $ is denoted by a dot. The line
element of $\Sigma _{\xi }$ in this coordinates is expressed by

\begin{equation}
ds_{s}^{2}=-d\tau ^{2}+R^{2}(\tau )d\Omega _{d-2}^{2}\text{.}  \label{dsp}
\end{equation}

The interior and exterior line element with spherical symmetry are
given by

\begin{equation}
ds_{-}^{2}=-f_{-}^{2}(r)dt_{-}^{2}+f_{-}^{-2}(r)dr^{2}+r^{2}d\Omega
_{d-2}^{2}\mbox{, }r<R(\tau )\text{,}  \label{ds-}
\end{equation}
and

\begin{equation}
ds_{+}^{2}=-f_{+}^{2}(r)dt_{+}^{2}+f_{+}^{-2}(r)dr^{2}+r^{2}d\Omega
_{d-2}^{2}\mbox{, }r> R(\tau )\text{.}  \label{ds+}
\end{equation}
Interior and exterior coordinates match continuously on the
$\Sigma _{\xi }$, but it is found that $t_{-}\neq t_{+}$. In these
coordinates the vectors $u^{\alpha } $ and $\xi ^{\alpha }$ are
given by

\begin{eqnarray}
u^{\alpha } &=&(\frac{\gamma }{f^{2}},\dot{R},0,...,0)\text{,}  \label{vel}
\\
\xi ^{\alpha } &=&(\frac{\dot{R}}{f^{2}},\gamma ,0,...,0)\text{,}
\label{normal}
\end{eqnarray}
where
\begin{equation}
\gamma =\sqrt{f^{2}+\dot{R}^{2}}\text{.}
\label{gammadef}
\end{equation}
Substituting (\ref{ds-}) and (\ref{ds+}) into the Hamiltonian
generator $\mathcal{H}_{\bot }$ (\ref{hp}), we obtain

\begin{equation}
\mathcal{H}_{\perp }=-\frac{\kappa (d-2)!}{r^{d-2}}\sqrt{g}\frac{d}{dr}%
\left\{ r^{d-1}\sum_{p}(d-2p)\alpha _{p}\left( \frac{1-f^{2}}{r^{2}}\right)
^{p}\right\} +\sqrt{g}T_{\perp \perp }\text{.}  \label{s4}
\end{equation}

We are interested in integrating out the constraint
$\mathcal{H}_{\bot }=0$ across a radial infinitesimal length
centered in the shell position $r= R(\tau )$ to a constant time.
In this form, it is possible to express the discontinuities of
geometry in terms of the projected stress $T_{\perp \perp }$. It
is easy to prove that $T_{\perp \perp }$ has the same form that
the one obtained in \cite{cot}, due to the symmetry of the thin
shell. Finally $T_{\perp \perp }$ is given by

\begin{equation}
T_{\perp \perp }=\sigma \gamma \delta \left( r-R(\tau )\right) \text{.}
\label{s5}
\end{equation}
Integrating in the radial direction, from $R+\epsilon$ and
$R-\epsilon$, we obtain in the limit $\epsilon \rightarrow 0$

\begin{equation}
\kappa (d-2)!R\sum_{p}(d-2p)\alpha _{p}\left\{ \left( \frac{1-f_{+}^{2}(R)}{%
R^{2}}\right) ^{p}-\left( \frac{1-f_{-}^{2}(R)}{R^{2}}\right) ^{p}\right\} =%
\frac{1}{2}\sigma (\gamma _{+}+\gamma _{-})\text{.}
\label{juancho}
\end{equation}
This expression has been seen as the generalization of the
equation obtained for \textbf{GR}.

From expression (\ref{she1}), it is found that

\begin{equation}
M_{+}-M_{-}=\frac{1}{2}m\left( \gamma _{+}+\gamma _{-}\right) \text{,}
\label{s6}
\end{equation}
where $m=\Omega _{d-2}R^{d-2}\sigma $. Expression (\ref{s6}) is
the same to that obtained from the \textbf{GR} case \cite{IS}. If
$\sigma >0$ then $M_{+}>M_{-}$, this means that if $M_{-}$ is the
mass of a black hole inside of the shell, the final mass of the
black hole will be greater, therefore its events horizon
increases.

In order to complete the present picture of a radial collapse, it
is necessary to analyze the consistency of the remaining
nonvanishing components of the Hamiltonian treatment related to
the radial and angular components. The angular contribution of the
constraint (\ref{hi}) are identical to zero. Because the radial
component is not identically zero, it is necessary to evaluate
\begin{equation}
\mathcal{H}_{r}=-2\pi _{r/j}^{j}+2\sqrt{g}T_{\bot r} \text{,}
\label{rligadura}
\end{equation}
which yields
\begin{equation}
\mathcal{H}_{r}=\kappa \sqrt{\Gamma }(d-2)!\sum\limits_{p=0}^{k}\frac{%
p!(d-2p)!\alpha _{p}}{2^{p+1}(d-2p-1)!}\sum%
\limits_{s=0}^{p-1}2^{s}D_{s(p)}f^{2(s-p)}\left( 1-f^{2}\right) ^{s}\frac{d}{%
dr}\left( r^{d-2p-1}(\alpha \dot{r})^{2p-2s-1}\right) -\frac{R^{d-2}\sqrt{\Gamma }\dot{R}}{f^{2}}%
\sigma \delta (r-R(\tau ))\text{,} \label{rliga}
\end{equation}
where $\Gamma $ corresponds to the determinant of the angular
metric. One can expect that $\mathcal{H}_{r}$ to be proportional
to $\mathcal{H}_{\perp }$, since (\ref{hp}) already provides the
equation of motion for $R(\tau )$. It would be interesting to
explicitly see that this indeed occurs. But due to equation
(\ref{rliga}), in the general case it is complicated (it is not
possible to integrate) to perform this point. However, it is
straightforward to prove that the correct Einstein-Hilbert limit
is obtained when $k=1$ \cite{cot}.

The acceleration of the thin shell is obtained from equation
(\ref{s6}) by differentiating with respect to proper time $\tau$,
which yields

\begin{equation}
m\ddot{R}=-\frac{m^{2}}{2\left( M_{+}-M_{-}\right) }\left( \gamma _{+}\frac{%
df_{-}^{2}}{dR}+\gamma _{-}\frac{df_{+}^{2}}{dR}\right) -\left( d-2\right)
\Omega _{d-2}R^{d-3}\hat{\tau}\gamma _{+}\gamma _{-}\text{,}  \label{s7}
\end{equation}
Notice that we need the explicit forms of $f_{-}^{2}$ and $%
f_{+}^{2}$. The form of $f_{-}^{2}$ will be

\smallskip

\begin{equation}
f_{-}^{2}(r)=\frac{r^{2}}{l^{2}}+1\text{.}  \label{s9}
\end{equation}
The form of $f_{+}^{2}$ must be split  into two cases; the case
$d-2k-1\neq 0$ and the case $ d-2k-1=0$. In the latter case there
exists a gap in the mass, in which the vacuum corresponds to
$M_{-}=-(2G_{k})^{-1}$.

\begin{itemize}
\item  If $d-2k-1\neq 0$, with $M_{-}=0$ and $M_{+}=M$, then
equation (\ref{s6}) takes the form $M=\frac{1}{2}m\left( \gamma
_{+}+\gamma _{-}\right) $, so that if $\sigma >0$, then $M>0$. The
acceleration is given by

\begin{eqnarray}
m\ddot{R} &=&-\frac{m}{l^{2}}R-\left( d-2\right) \Omega _{d-2}R^{d-3}\hat{%
\tau}\gamma _{+}\gamma _{-}  \nonumber \\
&&-\frac{\chi \left( d-2k-1\right) m^{2}}{2k}\left( \frac{2G_{k}}{%
M^{k-1}R^{d-k-1}}\right) ^{1/k}\gamma _{-}  \label{accel}
\end{eqnarray}
The first two terms of (\ref{accel}) correspond to the
acceleration due to AdS, and the interaction of tangent tension on
the thin shell. If $\chi=1$, for a black hole solution, then
$\ddot{R}<0$. Therefore, in this way the thin shell always
collapses to a black hole.

 On the other hand, when
$\chi =-1$ we might think that a naked singularity , however due
to that the latter term a Eq. (\ref{accel}) is positive and thus a
repulsive gravitational force appears. It could be shown that this
force therefore domines over the other terms when $R\rightarrow0$.
Therefore, naked singularity cannot be formed through a thin shell
collapse. In order to see this point we consider

\begin{equation}
M=\frac{m}{2}\left(\gamma_{+}+\gamma_{-}\right),\label{M}
\end{equation}
where
\begin{equation}
\gamma_{\pm}=\sqrt{\dot{R}^{2}+f_{\pm}^{2}}, \label{gamma}
\end{equation}
and $m=\Omega_{d-2}R^{d-2}\sigma$ with $M>0$. In this case
$f_{-}^{2}$ and $f_{+}^{2}$ are given by
\begin{equation}
f_{-}^{2} =1+\frac{R^{2}}{l^{2}},
\end{equation}
and
\begin{equation}
 f_{+}^{2} =1+\frac{R^{2}}{l^{2}}+\left(
\frac{2G_{k}M}{R^{d-2k-1}}\right) ^{1/k},
\end{equation}respectively. For a naked singularity it is necessary that
$f^{2}_{+}>f^{2}_{-}>0$.

From (\ref{M}) we obtain $\dot{R}^{2}$,
\begin{equation}
\dot{R}^{2}=\left[\frac{M}{m}-m\left(\frac{2^{1-2k}G_{k}M^{1-k}}{R^{d-2k-1}}\right)^{1/k}\right]^{2}-
\left(\frac{R^{2}}{l^{2}}+1\right),\label{efectivo}
\end{equation}
from which we could read an effective potential.
\begin{figure}[th]
\includegraphics[width=5.0in,angle=0,clip=true]{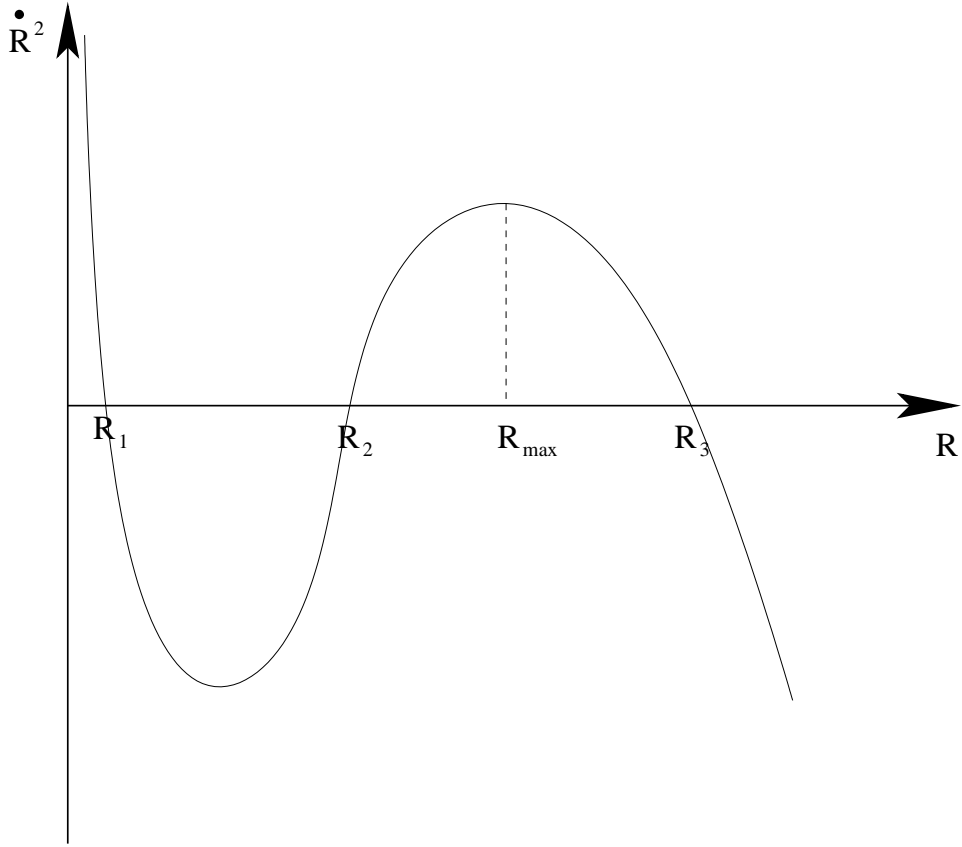}
\caption{This plot shows  behavior of our effective potential and
their turning point.} \label{fig1}
\end{figure}

 In order to give either a quantitative and qualitative
discussion of this potential, let us write Eq. (\ref{efectivo}) in
the form
\begin{equation}
\dot{R}^{2}=\alpha_{+}\alpha_{-}\label{chi}
\end{equation}
where
\begin{equation}
\alpha_{\pm}=\frac{M}{m}-m\left(\frac{2^{1-2k}G_{k}M^{1-k}}{R^{d-2k-1}}\right)^{1/k}\pm\sqrt{\frac{R^{2}}{l^{2}}+1}.
\end{equation}
The physical regions are defined by $\dot{R}^{2}>0$, which implies
that both $\alpha_{+}$ and $\alpha_{-}$ have the same sign.  Note
that for the case under study we have $\gamma_{\pm}>1$, due to
$f^{2}_{\pm}>1$, this means that the sum
$\gamma_{+}+\gamma_{-}>2$. Therefore, from Eq.(\ref{M}) we obtain
$M>m$. In order to characterize the physical regions, we need the
behaviors of the $\alpha_{\pm}$ parameters. We will simplify our
study to the dust case which means $m=Constant$. Our results are
shown in Fig.1 from which we could see different regions,

\begin{eqnarray*}
\mbox{I. }\dot{R}^{2} &>&0\mbox{, for }R<R_{1}\mbox{,} \\
\mbox{II. }\dot{R}^{2} &<&0\mbox{, for }R_{1}<R<R_{2}\mbox{,} \\
\mbox{III. }\dot{R}^{2} &>&0\mbox{, for }R_{2}<R<R_{3}\mbox{,} \\
\mbox{IV. }\dot{R}^{2} &<&0\mbox{, for }R_{3}<R\mbox{,}
\end{eqnarray*}

Note that the classical allowed regions are I and III. In region I
note that when $R\rightarrow 0$ $\dot{R}^{2}\rightarrow\infty$. In
this case, a naked singularity emerge from a collapse of a thin
shell. However we could prove that the motion in this region is
prohibited by causality. In fact,  from Eq. (\ref{gama}) ( see
appendix), we obtain
\begin{equation}
\gamma_{-}=\frac{M}{m}-m\left(\frac{2^{1-2k}G_{k}M^{1-k}}{R^{d-2k-1}}\right)^{1/k}\label{alfa}
\end{equation}
and the condition $\gamma_{-}>0$, implies
\begin{equation}
\frac{M}{m}>m\left(\frac{2^{1-2k}G_{k}M^{1-k}}{R^{d-2k-1}}\right)^{1/k}.\label{rmin}
\end{equation}
This mean that it  must exist a minimum radius, which we denote by
$R=R_{*}$ in order not to violate the conservation law expressed
by Eq. (\ref{M}). It is direct to prove that $R_{1}<R_{*}<R_{2}$,
which  implies that $\alpha_{+}>0$. Therefore, the thin shell
"movement" is allowed only  in region III, this is
$R_{2}<R<R_{3}$.  So, naked singularity, in the case of dust,
cannot be formed through a thin shell collapse.

For the general case, is not simple to study equation
(\ref{efectivo}), since it requires to solve equation
(\ref{state}) (see appendix). Also we should know the relation
between $\sigma$ and $\hat{\tau}$ (the state equation for the
shell), in order to obtain $m=m(R)$. However, we may argue that
equation (\ref{alfa}) is of general character, so also it applies
for the general case. The a minimum radius must exist, so that the
thin shell does not violate equation (\ref{alfa}), Therefore, it
must exist a turning point. So, naked singularity, in the general
case, cannot be formed through a thin shell collapse.

 It is easy to see that in the limit $R\rightarrow 0$ the first
term vanishes. If we consider the particular case of dust, i.e.
$\tilde{\tau} = 0$, eq. (\ref{state}) implies that $m = Const.$,
therefore the last term goes to infinity for $R \rightarrow 0$.
Thus, for the dust case the thin shell does not collapse to $R =
0$, because the acceleration becomes very strong.

At this point, if we considered Einstein-Hilbert limit ($k=1$) in
equation (\ref{accel}), the last term  would be reduced to
-$(d-3)Gm^{2}/R^{d-2}$, thus corresponding to the Newton
gravitational interaction. Therefore,for $k\neq 1$ this term will
be a generalization of the Newton gravitational interaction, with
an effective gravitational constant given by
\begin{equation}
-\frac{1}{k}\left( \frac{2G_{k}}{%
M^{k-1}}\right) ^{1/k}
\end{equation}

\item  If $d-2k-1=0$, it corresponds to the {\bf{CS}} theory with
$M_{-}=-(2G_{k})^{-1}$. In this case, expression (\ref{s6}) takes
the form $M+(2G_{k})^{-1}=\frac{1}{2}m\left( \gamma _{+}+\gamma
_{-}\right) $. If $\sigma >0$, then it is implied that
$M>-(2G_{k})^{-1}$; therefore, the naked singularities with
negative mass can emerge from the collapse of a thin shell. Here,
acceleration is given by

\begin{equation}
m\ddot{R}=-\frac{m}{l^{2}}R-\left( d-2\right) \Omega _{d-2}R^{d-3}\hat{\tau}%
\gamma _{+}\gamma _{-}\text{,}  \label{s12}
\end{equation}

where $\ddot{R}<0$, and thus the thin shell always collapses. It
is observed from eq. (\ref{s12}) that a term analogue to the
Newton gravitational interaction does not appear.

\end{itemize}

\section{Conclusion and Remarks}
We have developed the Hamiltonian formalism for the collapse of
thin shells in Lanczos-Lovelock theories, and we presented given a
concise analysis of the theories described in Ref. \cite{ctz}. We
show in these theories that the black holes solution can be
created by collapsing a thin shell and naked singularities cannot
be formed by this mechanism. On the other hand, if we consider
theories with $k\neq 1$, these exhibit a generalization of the
Newton gravitational interaction, and effective gravitational
constant becomes given by
\begin{equation}
-\frac{1}{k}\left( \frac{2G_{k}}{ M^{k-1}}\right) ^{1/k}.
\end{equation}
Also we have shown that when we take the Einstein-Hilbert limit
($k=1$) in equation (\ref{accel}), the last term is reduced to
-$(d-3)Gm^{2}/R^{d-2}$,  which coincides with the Newton
gravitational interaction.

Nevertheless among the different Lanczos-Lovelock's theories,  the
Chern-Simons theory exhibits an exceptional behavior, since naked
singularities can emerge from the collapsing of a thin shell. This
kind of theory does not possess a gravitational self-interaction
analogous to the Newtonian case.

It is straightforward to prove that in the case of electrically
charged thin shells, the general form of eq. (\ref{juancho}),
governing the radial collapse in $d$ dimensions, remains the same
because the electromagnetic stress tensor contributes with a
finite jump value on the $\mathcal{H}_{\bot }$ and, therefore,
does not contribute to the radial integral of $\mathcal{H}_{\bot
}$. Moreover, when we consider the charged case in \textbf{CS}
theories, a mechanism that prevents the naked singularities
formation appears.

Finally, we conjecture, by virtue of the results from Ref.
\cite{cot} that the presence of an angular moment in $2+1$
dimensions prevents the formation of naked singularities; thus, in
higher dimensions, the angular moment could prevent naked
singularities formations in \textbf{CS} theories.

\section{Acknowledgments}
We are grateful to S. Lepe, R. Olea, R. Troncoso, C. Teitelboim
and J. Zanelli for many enlightening discussions. S.d.C. was
supported from COMISION NACIONAL DE CIENCIAS Y TECNOLOGIA through
FONDECYT Grant Nos. 1030469; 1010485 and 1040624. Also, it was
partially supported by UCV  Grant No. 123.764/2004. J.S. was
supported from COMISION NACIONAL DE CIENCIAS Y TECNOLOGIA through
FONDECYT Postdoctoral Grant 3030025.  JC was supported by
Ministerio de Educaci\'{o}n through a MECESUP grant, FSM 9901. The
authors wish to thank CECS for its kind hospitality.

\section{Appendix}

From Eq. (\ref{efectivo}), \textit{i.e.}
\begin{equation}
\dot{R}^{2}=\left[\frac{M}{m}-m\left(\frac{2^{1-2k}G_{k}M^{1-k}}{R^{d-2k-1}}\right)^{1/k}\right]^{2}-
\left(\frac{R^{2}}{l^{2}}+1\right)\mbox{,}\label{efectivo1}
\end{equation}
we will obtain the acceleration of the thin shell given by Eq.
(\ref{accel}).

Let star from
\begin{equation}
\gamma_{\pm}=\frac{M}{m}\pm\left(\frac{2^{1-2k}G_{k}M^{1-k}}{R^{d-2k-1}}\right)^{1/k}m\mbox{,}\label{gama}
\end{equation}
and
\begin{equation}
(\sigma u^{a})_{/a}-\hat{\tau}u^{a}_{/a}=0\mbox{.}\label{state}
\end{equation}
This equation can be seen as the equation of state of the matter
on the hypersurface $\Sigma _{\xi }$. Using spherical coordinates
and the identity
$T^{a}_{/a}=\partial_{a}(\sqrt{-h}T^{a})/\sqrt{-h}$, where $h$ is
the proper metric determinant, we obtain
\begin{equation}
\partial_{a}(\sqrt{-h}\sigma
u^{a})=\hat{\tau}\partial_{a}(\sqrt{-h}u^{a})\mbox{.}\label{estateesfe}
\end{equation}
The proper metric is given by
\begin{equation}
ds^{2}=-d\tau ^{2}+R^{2}(\tau )d\Omega _{d-2}^{2}\mbox{,}
\label{dsp}
\end{equation}
with $\sqrt{-h}=R^{d-2}\sqrt{\Gamma}$, where $\Gamma$ is the
angular metric determinant. Besides we consider radial collapse
then $u^{a}=(1,0,...,0)$, moreover consider proper coordinates
given by $\rho^{a}=(\tau,\theta^{1},...,\theta^{d-2})$, then we
get
\begin{equation}
\frac{d}{d\tau}\left(R^{d-2}\sqrt{\Gamma}\sigma\right)=(d-2)R^{d-3}\hat{\tau}\sqrt{\Gamma}\dot{R}
\mbox{,}
\end{equation}
where $\dot{R}=dR/d\tau$ and since $d\sqrt{\Gamma}/d\tau=0$ we
have
\begin{equation}
\frac{d}{d\tau}\left(R^{d-2}\sigma\right)=(d-2)R^{d-3}\hat{\tau}\dot{R}\mbox{.}\label{state}
\end{equation}
Multiplying this latter equation by the angular volume of the unit
sphere, $\Omega_{d-2}$ (for $d=4$  $\Omega_{2}=4\pi$) and defining
$m=\Omega_{d-2}R^{d-2}\sigma$, we obtain
\begin{equation}
\frac{dm}{d\tau}=(d-2)R^{d-3}\hat{\tau}\dot{R}\mbox{.}
\end{equation}
Due to, symmetry we have $m=m(R)$ and $d/dt=\dot{R}d/dR$, then
\begin{equation}
\frac{dm}{dR}=(d-2)R^{d-3}\hat{\tau}\mbox{.}\label{mr}
\end{equation}

From which we get
\begin{eqnarray}
2\ddot{R}\dot{R} &=&-\frac{2}{l^{2}}R\dot{R}\nonumber \\
&&
+2\left[\frac{M}{m}-m\left(\frac{2^{1-2k}G_{k}M^{1-k}}{R^{d-2k-1}}\right)^{1/k}\right]\nonumber\\
&&\left[-\frac{M}{m^{2}}\frac{dm}{d\tau}-\left(\frac{2^{1-2k}G_{k}M^{1-k}}{R^{d-2k-1}}\right)^{1/k}
\frac{dm}{d\tau}+m\frac{(d-2k-1)}{k}\left(\frac{2^{1-2k}G_{k}M^{1-k}}{R^{d-k-1}}\right)^{1/k}\dot{R}\right].\label{accel1}
\end{eqnarray}
For the $(-)$ sign, we obtain the following
\begin{eqnarray}
2\ddot{R}\dot{R} &=&-\frac{2}{l^{2}}R\dot{R}\nonumber \\
&& +2\gamma_{-}
\left[-\frac{M}{m^{2}}\frac{dm}{d\tau}-\left(\frac{2^{1-2k}G_{k}M^{1-k}}{R^{d-2k-1}}\right)^{1/k}
\frac{dm}{d\tau}+m\frac{(d-2k-1)}{k}\left(\frac{2^{1-2k}G_{k}M^{1-k}}{R^{d-k-1}}\right)^{1/k}\dot{R}\right]\mbox{.}\label{accel2}
\end{eqnarray}
we could rewritten this equation in the form
\begin{eqnarray}
\ddot{R}\dot{R} &=&-\frac{1}{l^{2}}R\dot{R}\nonumber \\
&& -\frac{\gamma_{-}}{m}
\left[\frac{M}{m}+\left(\frac{2^{1-2k}G_{k}M^{1-k}}{R^{d-2k-1}}\right)^{1/k}m
\right]\frac{dm}{d\tau}+\frac{(d-2k-1)}{k}\left(\frac{2^{1-2k}G_{k}M^{1-k}}{R^{d-k-1}}\right)^{1/k}m\dot{R}\gamma_{-}
\mbox{.} \label{accel3}
\end{eqnarray}
For $(+)$ sign, we obtain
\begin{eqnarray}
\ddot{R}\dot{R}
&=&-\frac{1}{l^{2}}R\dot{R}-\frac{\gamma_{-}\gamma_{+}}{m}\frac{dm}{d\tau}\nonumber\\
&&+\frac{(d-2k-1)}{k}\left(\frac{2^{1-2k}G_{k}M^{1-k}}{R^{d-k-1}}\right)^{1/k}m\dot{R}\gamma_{-}\mbox{.}
\label{accel4}
\end{eqnarray}
but $dm/d\tau=\dot{R}dm/dR$ and using Eq. (\ref{mr})
\begin{eqnarray}
\ddot{R}
&=&-\frac{1}{l^{2}}R-(d-2)R^{d-3}\hat{\tau}\frac{\gamma_{-}\gamma_{+}}{m}\nonumber\\
&&+\frac{(d-2k-1)}{k}\left(\frac{2^{1-2k}G_{k}M^{1-k}}{R^{d-k-1}}\right)^{1/k}m\gamma_{-}\mbox{,}
\label{accel5}
\end{eqnarray}
multiplying by $m$, we obtain finally
\begin{eqnarray}
m\ddot{R}
&=&-\frac{m}{l^{2}}R-(d-2)R^{d-3}\hat{\tau}\gamma_{-}\gamma_{+}\nonumber\\
&&+\frac{(d-2k-1)}{k}\left(\frac{2^{1-2k}G_{k}M^{1-k}}{R^{d-k-1}}\right)^{1/k}m^{2}\gamma_{-}\mbox{,}
\label{accel6}
\end{eqnarray}
which corresponds to Eq. (\ref{accel}) in the main text.

\end{document}